%% file: main.tex
\documentclass[oneside,letterpaper,10pt,onecolumn]{article}
\input{mypackages.tex}

\input{mymath.tex}

\usepackage[numbers]{natbib}
\usepackage{mathtools}
\usepackage{algorithmic}
\usepackage[ruled,vlined,boxed,linesnumbered]{algorithm2e}
\DeclarePairedDelimiterX{\infdivx}[2]{(}{)}{%
  #1\;\delimsize\|\;#2%
}

\title{Online Influence Maximization: Concept and Algorithm}
\author{\textbf{Jianxiong Guo}\\ 
Advanced Institute of Natural Sciences,\\ 
Beijing Normal University, Zhuhai 519087, China \\
jianxiongguo@bnu.edu.cn}

\date{}

\begin{document}
\maketitle

\section{Introduction}
Influence Maximization (IM) is a fundamental problem in the field of network analysis, particularly relevant in the domains of social networks, marketing, and information dissemination. At its core, the problem revolves around identifying the most influential nodes (individuals, entities, or elements) within a network to maximize the spread of information, behavior, or trends. Kempe et al. \cite{kempe2003maximizing} first defined the IM problem as a combinatorial problem that selects a small subset of nodes such that maximizes the expected number of activated nodes (influence spread). This concept has gained significant attention with the rise of online social networks, where the spread of information can have far-reaching impacts. The influence spread is often modeled using various diffusion models, such as Independent Cascade (IC) and Linear Threshold (LT) models, which simulate how information or influence propagates through the network. Thus, the IM problem is usually characterized by a directed graph and a diffusion model together with model parameters (graph weights), which has been widely used in many real-world applications \cite{li2018influence}.

Diffusion models, such as IC and LT models, rely on parameters that define how influence propagates from one node to another in a network. These parameters might include the probability of a node influencing its neighbor, the threshold of influence required for a node to be activated, or the overall strength of influence between nodes. However, in real-world applications, these parameters are rarely directly observable and can be highly complex, which are usually unknown. A heuristic method to overcome this issue is to estimate the unknown parameters with high confidence and low error from the collected past observations, which is usually called the ``Network Inference'' problem \cite{goyal2010learning}\cite{du2014influence}\cite{narasimhan2015learnability}\cite{kalimeris2018learning}\cite{conitzer2020learning}\cite{chen2021network}\cite{conitzer2022learning}\cite{huang2023multi}. However, they may face the difficulties of insufficient data logs, biases in data logs, and inability to adapt to changes. Here, we summarize the challenges posed by unknown parameters in the IM problem as follows: (1) \textbf{Complex Social Dynamics:} Social networks are characterized by complex and dynamic interactions. The factors influencing how information spreads can vary widely depending on the context, the nature of the information, and the characteristics of individuals involved. (2) \textbf{Data Scarcity and Quality:} Obtaining accurate and comprehensive data to estimate these parameters is challenging. Data might be incomplete, noisy, or biased, leading to inaccurate estimations of diffusion probabilities. (3) \textbf{Adaptive Behaviors:} Individuals in a network may change their behavior over time, influenced by external factors or their interactions within the network. This adaptive behavior makes it difficult to maintain a consistent set of parameters for influence diffusion. (4) \textbf{Diversity of Influence:} Different nodes may have different levels and types of influence. A one-size-fits-all approach to parameter estimation often fails to capture the nuances of individual nodes' influence.

Instead of network inference, the Online IM problem \cite{lei2015online} represents a strategic shift in addressing the challenges of IM in dynamic and uncertain environments, like social networks. Unlike traditional IM approaches that rely on pre-existing models with fixed parameters, Online IM adopts an iterative, learning-based approach. This method allows for the progressive refinement of diffusion models and the identification of optimal seed sets based on real-time interactions with the network. The Online IM problem can be characterized by the following six points: (1) \textbf{Iterative Learning:} It operates through a series of rounds. In each round, a set of seed nodes is selected based on the current understanding of the network's influence dynamics. The outcomes of these selections (i.e., how far and effectively the influence spreads from these seeds) are then observed and recorded. (2) \textbf{Parameter Estimation and Update:} The observed outcomes are used to update the model's understanding of the network's influence parameters. This could involve adjusting the probabilities of influence between nodes, the thresholds for activation, or other model-specific parameters. The key is that these updates are based on actual observed data from the network, making them more accurate and relevant. (3) \textbf{Progressive Optimization:} With each iteration, the model becomes more refined and better at predicting which nodes will be most effective at spreading influence. This iterative refinement helps in progressively moving towards the identification of an optimal seed set – a set of nodes that can maximize influence spread within the network. (4) \textbf{Adaptation to Network Dynamics:} One of the significant advantages is its inherent adaptability. Because it continuously updates its parameters based on ongoing interactions, it can adapt to changes in the network's structure or in the behavior of its nodes. This dynamic adaptability is crucial in the ever-changing landscape of social networks. (5) \textbf{Balancing Exploration and Exploitation:} It typically involves a balance between exploration (trying out different seed sets to learn more about the network's influence dynamics) and exploitation (using the best-known strategies to maximize influence spread). Finding the right balance is key to the efficiency and effectiveness of strategies. (6) \textbf{Computational Considerations:} While it provides a more adaptive and potentially more accurate approach to IM, it also comes with computational challenges. Each iteration requires data collection, processing, and model updating, which can be resource-intensive. Therefore, efficient algorithms and scalable computing resources are often necessary for practical Online IM applications.

In summary, the Online IM problem offers a dynamic and adaptive approach to identifying influential nodes in social networks. By learning from real-time interactions and continuously refining its strategies, it can effectively navigate the uncertainties and complexities of real-world social networks, making it a powerful tool for influence maximization in various applications. Therefore, in this paper, we aim to make a comprehensive overview of the Online IM problem by considering all recent developments in this area. In Section \ref{sec2}, we review the basic definitions and algorithms of the Offline IM problem because they will be invoked by Online IM algorithms. In Section \ref{sec3}, we systematically introduce the ins and outs of the Online IM problem and discuss the most important theoretical foundation to study the Online IM problem: Combinatorial Multi-Armed Bandit (CMAB). In Section \ref{sec4}, we divide Online IM algorithms into three categories, edge-level feedback, node-level feedback, and other feedback, and introduce the main work of each category in detail. In Section \ref{sec5}, we consider some non-standard research works about the IM problem and variants in real applications, which greatly extends the research method and application scope of the Online IM problem. In Section \ref{sec6} and \ref{sec7}, we summarize several future research and development directions and conclude this paper.

\section{Preliminaries: Offline Influence Maximization}\label{sec2}
In this section, we first introduce the preliminary of social networks and then give the standard Offline Influence Maximization (IM) problem. Finally, we summarize the mainstream methods and techniques to address the IM problem.

\subsection{Background of Social Networks}
For a given social network, we can represent it as a directed graph $G=(V,E)$, where the set of nodes $V=\{v_1,v_2,\cdots,v_{|V|}\}$ and the set of edges $E=\{e_1,e_2,\cdots,e_{|E|}\}$ are defined. Each node $v$ in $V$ signifies a user, and each edge $e=(u,v)$ in $E$ symbolizes a certain type of connection from user $u$ to $v$. For each edge $(u,v)\in E$, $u$ is termed as the in-neighbor of $v$, and conversely, $v$ is the out-neighbor of $u$. For any given node $v$ in $V$, the collection of its in-neighbors is denoted as $N^-(v)$, and the collection of its out-neighbors is denoted as $N^+(v)$.

In the process of information diffusion, a user is deemed active when she accepts (is activated by) the information from her in-neighbors or is chosen as an initial influencer (seed). This information cascade is often characterized by a diffusion model, for instance, the Independent Cascade (IC) model \cite{kempe2003maximizing}.
\begin{defn}[IC model]
    Under the IC model, with an initial seed set $S\subseteq V$, the process of stochastic information spread is as follows: (1) Initially, at timestamp $0$, all nodes in $S$ are in an active state, while the rest of the nodes in $V\backslash S$ remain inactive. Once activated, a node stays active; (2) At any timestamp $\tau$, every newly activated node $u$ gets a single opportunity to switch its inactive out-neighbor $v$ to an active state with a probability of $p_{uv}$ at timestamp $\tau+1$; and (3) The process of information spreading concludes when there are no further inactive nodes that can be activated in the upcoming timestamps.
\end{defn}
We believe that a diffusion model can be uniquely determined by its diffusion mechanism and parameters, thus when the diffusion mechanism is assumed in advance, learning the diffusion process of the diffusion model is equivalent to learning its parameters. This rule is very important when discussing online problems later. Subsequently, we take the IC model $\mathcal{D}(G,\vp)$  as an example to discuss our problem, which can be parameterized by a parameter set $\vp=\{p_{uv}:(u,v)\in E\}$ given a social graph $G$. (When the context is clear, the IC model can be abbreviated as $\mathcal{D}$.) This is because most current investigations about the Online IM problem are based on the IC model because of its simplicity. \textit{When there is no special explanation, we can default that the potential diffusion model is the IC model, of course, our discussion will also involve other diffusion models, such as Linear Threshold (LT) and Triggering (TR) models \cite{kempe2003maximizing}.}

\subsection{Influence Maximization}
The Offline IM problem involves finding a small subset of nodes (seed set) $S\subseteq V$ such that its influence spread $\sigma_{\mathcal{D}}(S)$ can be maximized under the diffusion model $\mathcal{D}$, which is the expected number of active nodes after the diffusion terminates. 

To formally describe the influence spread, we first need to understand the notion of ``realization''. A realization, denoted as $g = (V, E_g)$ where $E_g\subseteq E$, represents a subgraph that is randomly generated based on the diffusion model. Taking the IC model as an instance, each edge $(u, v)\in E$ is independently included in $E_g$ with a likelihood of $p_{uv}$. Edges contained in $E_g$ in the realization $g$ are termed ``live'' edges. Thus, the probability of realization $g$ sampled based on the IC model $\mathcal{D}(G,\vp)$ is $\Pr[g;\mathcal{D}(G,\vp)]=\prod_{e\in E_g}p_e\prod_{e\in E\backslash E_g}(1-p_e)$. Considering there are $2^{|E|}$ potential realizations, the influence cascade in any realization is deterministic, not stochastic. Therefore, the influence spread through the network is essentially the average spread over all these possible realizations. The Offline IM problem is thus expressed in terms of expected values and can be formally defined accordingly.
\begin{defn}[Influence Maximization]
    In the context of a social graph $G=(V, E)$, considering a diffusion model (specifically, the IC model discussed in this paper), and given a limit of $k$ nodes (referred to as the budget), the IM problem is to identify an optimal seed set $S^\circ$ with at most $k$ nodes that is intended to maximize the expected influence spread throughout the network, i.e.
    \begin{equation}
        S^\circ\in\arg\max_{|S|\leq k}\sigma_{\mathcal{D}}(S)=\mathbb{E}_{g\sim\mathcal{D}(G,\vp)}[|I_g(S)|]=\sum\nolimits_{g\in\mathcal{G}}\Pr[g;\mathcal{D}(G,\vp)]\cdot |I_g(S)|,
    \end{equation}
    where $\mathcal{G}$ is the ensemble of all conceivable realizations and $I_g(S)$ is the set of nodes that encompasses all nodes that are reachable from any node in $S$ through the live edges in the realization $g$.
\end{defn}

The influence function $\sigma$ is a set function. For any set function $f : 2^V\rightarrow\mathbb{R}_+$ and any two subsets $S$ and $T$ where $S\subseteq T\subseteq V$, it is monotone if $f(S)\leq f(T)$ and submodular if $f(S\cup\{v\})-f(S)\geq f(T\cup\{v\})-f(T)$ for any $v\in V\setminus T$. The challenge of the IM problem is NP-hard nature, and under both IC and LT modes, the influence function is monotone and submodular. When we apply a cardinality constraint $|S|\leq k$, a greedy algorithm that employs a hill-climbing approach by consistently selecting the node with the highest marginal gain approximates the optimal solution to within a factor of $(1-1/e)$ \cite{nemhauser1978analysis}. However, calculating $\sigma(S)$ for a given seed set $S$ is \#P-hard under the IC \cite{chen2010scalable} and LT models \cite{chen2010scalable11}. The most straightforward method to estimate the influence spread is through Monte Carlo (MC) simulations, which involve repeated simulations of the diffusion process for a specific seed set and averaging the outcomes. This method introduces an error term $\varepsilon$, and as a result, the greedy algorithm, under the MC simulation framework, can only achieve a $(1-1/e-\varepsilon)$ approximation, with a running time of $\Omega(k|V||E|\cdot poly(1/\varepsilon))$ \cite{kempe2003maximizing}.

\subsection{Summarization of IM Algorithms}\label{sec2-3}
In the implementation of the greedy algorithm, MC simulations are utilized to approximate the marginal gains of all nodes during each iteration. This approach, however, results in prohibitively long computation times, especially in extensive social networks. Consequently, numerous subsequent studies have concentrated on enhancing the operational efficiency of IM algorithms. The primary objective of these efforts is to accurately estimate the influence spread, thereby addressing the challenges posed by the \#P-hard nature of the problem. With the development of artificial intelligence, a new method of solving combinatorial optimization problems using Machine Learning (ML) technologies, such as Graph Neural Networks (GNNs) and Deep Reinforcement Learning (DRL) has emerged. Based on that, IM algorithms can be roughly categorized into two big families, Traditional IM algorithms \cite{li2018influence} and ML-based IM algorithms \cite{li2023survey}.

\textbf{Traditional IM Algorithms.} 
Based on the theoretical worst-case guarantees, conventional algorithms for the IM problem are generally classified into two types: heuristic algorithms and approximation algorithms. Furthermore, depending on the various methodologies employed for estimating the influence spread, IM algorithms can be grouped into three distinct categories: simulation-based, proxy-based, and sampling-based approaches \cite{li2018influence}. Typically, most simulation-based and sampling-based approaches are developed with theoretical guarantees in mind. In contrast, proxy-based approaches that employ proxy-based strategies are often heuristic in nature. Each of these approaches will be elaborated as follows.
\begin{itemize}
    \item \textbf{Simulation-Based Approach:} It refers to using MC simulations to estimate the influence spread. The seminal work of the IM problem \cite{kempe2003maximizing} has used this method to implement influence maximization. The inefficiency in runtime of the simulation-based greedy algorithm has sparked considerable research focused on enhancing its efficiency. One direction is to reduce the number of MC simulations in each estimation by exploiting the submodular trait of the influence spread, such as CELF/CEFL++ \cite{leskovec2007cost}\cite{goyal2011celf++} and UBLF \cite{zhou2015upper}. An alternative approach involves diminishing the MC complexity by implementing heuristic rules that streamline MC simulations, such as CGA \cite{wang2010community} and SA \cite{jiang2011simulated}.

    \item \textbf{Proxy-Based Approach:} It refers to adopting a proxy model instead of the real diffusion model to simplify the information diffusion process and estimate the influence spread. 
    Proxy models are typically developed based on heuristic principles. These principles include ``transforming the stochastic diffusion model into a deterministic one to enable precise computation of influence spread'' or ``limiting the influence range of each node to its immediate neighbors and disregarding the influence on nodes beyond that range''. Its representative work includes SP \cite{kimura2006tractable}, MIA/PMIA \cite{chen2010scalable}, and IPA \cite{kim2013scalable} for the IC model and LDGA \cite{chen2010scalable1}, SIMPATH \cite{goyal2011simpath}, and EASYIM \cite{galhotra2016holistic} for the LT model.

    \item \textbf{Sampling-Based Approach:} It refers to estimating the influence spread based on a collection of realizations that are sampled in advance given a diffusion model. Regarding the effectiveness and time efficiency, current researchers focus more on Reverse Influence Sampling (RIS) \cite{borgs2014maximizing} rather than Forward Influence Sampling (FIS) \cite{chen2009efficient} even though FIS-based algorithms appear earlier. The RIS-based algorithms first generate a collection of random Reverse Reachable (RR) sets and then transform the IM problem to the maximum coverage problem, and finally, the greedy algorithm is used to cover as many as RR sets. Its representative work includes TIM/TIM+ \cite{tang2014influence}, IMM \cite{tang2015influence}, SSA/D-SSA \cite{nguyen2016stop}, OPIM-C \cite{tang2018online}, and HIST \cite{guo2022influence}. They all guarantee to return a $(1-1/e-\varepsilon)$ approximation with at least $(1-1/|V|)$ probability within a time complexity close to linearity.
\end{itemize}

\textbf{ML-Based IM Algorithms.} The IM problem is a kind of difficult Combinatorial Optimization (CO) problem due to the \#P-hardness to calculate the influence spread. There is no perfect solution to solve the IM problem by using traditional algorithms. Heuristic algorithms are tailored to specific diffusion models and lack theoretical guarantees. Approximation algorithms have hit a scalability ceiling, making further optimization of their time efficiency challenging. Regardless of the traditional algorithm employed, they generally suffer from limited generalization capabilities. This means that any alteration in the network's topology necessitates re-running the algorithm to maintain the relevance of the seed set. In recent years, DRL-based methods have emerged to solve the graph-based CO problem, which can effectively overcome the drawbacks of traditional algorithms. According to different embedding strategies for representing real-time states, the DRL-based algorithms can be categorized into two classes \cite{li2023survey}: Ptr-Nets Based RL and GNNs Based RL Frameworks. We will describe them one by one below.
\begin{itemize}
    \item \textbf{Ptr-Nets Based RL Framework:} 
    Introduced in 2015, Pointer Networks (Prt-Nets) \cite{vinyals2015pointer} represent a groundbreaking approach designed to address three complex geometric CO problems: Planar Convex Hulls, Delaunay Triangulations, and Planar Traveling Salesman Problem (TSP). Drawing inspiration from the sequence-to-sequence (seq2seq) model used in machine translation, Prt-Nets tackle CO problems by encoding the input sequence of a problem instance through an encoder. It then utilizes a decoder and attention mechanism \cite{bahdanau2015neural} to calculate the selection probability for each node. However, this supervised training approach demands a substantial volume of labeled data and inherently restricts solution quality to the level of the training data. To overcome these limitations, Bello et al. \cite{bello2017neural}  applied a DRL framework to train Ptr-Nets on CO problems. This approach yielded near-optimal solutions for TSP and optimal solutions for the knapsack problem. Motivated by this, a lot of subsequent researchers extended Ptr-Nets based RL algorithms to deal with other graph-based NP-hard CO problems \cite{nazari2018reinforcement}\cite{deudon2018learning}\cite{ma2019combinatorial}.
    
    \item \textbf{GNNs Based RL Framework:} 
    GNNs leverage deep learning methodologies to efficiently process data characterized by graph structures, extracting profound insights from such graph-based data \cite{zhang2020deep}. The S2V-DQN model \cite{khalil2017learning}, a widely recognized baseline, employs structure2vec for encoding, transforming node information of a graph into a hidden node representation. In this model, the Q-network, integral to the Deep Q-learning (DQN) algorithm, uses the current partial solution combined with the latent node representation to estimate the Q-value for each new node. Following a greedy policy, the model sequentially adds the node with the highest Q-value to the evolving solution set. Motivated by this, a lot of subsequent researchers proposed improved GNNs based RL algorithms, such as GCN-TREESEARCH \cite{li2018combinatorial} and GCOMB \cite{manchanda2020gcomb}. It can adapt to solve different CO problems with different requirements and constraints in real applications, such as route optimization \cite{almasan2022deep}, scheduling \cite{mao2019learning}, and VNF placement \cite{sun2020combining}.
\end{itemize}

Recently, a series of works has transferred the above GNNs based RL framework to the IM problem. Fan et al. \cite{fan2020finding} used GraphSAGE as a function approximator in DQN to find key participants in complex networks. Li et al. \cite{li2022piano} proposed a PIANO model, which further expanded the S2V-DQN model, and used subgraphs to train DQN network to meet the requirements of solving the generalization in large-scale social networks. Ma et al. \cite{ma2022influence} modeled the IM problem as a continuous parameter optimization problem based on DQN network and proposed EDRL-IM model combined with an evolutionary algorithm to update DQN, which achieved good test results. Chen et al. \cite{chen2023touplegdd} put forward a ToupleGDD model by using three different groups of GNNs to encode the network state and learn the parameters of Double DQN, which is trained on small-scale random graphs, and then applied to completely different social graphs and achieves good generalization. 
Lin et al. \cite{ling2023deep} developed DeepIM, a model that generatively captures the latent representations of seed sets. In this work, they aim to comprehend and learn the varied patterns of information diffusion in a manner that is both data-driven and end-to-end.

\section{Online Influence Maximization}\label{sec3}
In the last section, we briefly introduce the IM problem, or called the Offline IM problem, in order to distinguish it from the Online IM problem. Regardless of traditional and ML-based IM algorithms, they are all based on a basic assumption that the diffusion model and its parameters are known in advance. For example in the IC model, the diffusion probability $p_{uv}$ of each edge $(u,v)\in E$ is assigned with a $1/|N^-(v)|$ or sampled from a given distribution, which is obviously not in line with reality and cannot describe the diversified information propagation process. Thus, a new research question was born: Online Influence Maximization.

The Online IM problem was first proposed in 2015 \cite{lei2015online}, where they laid the basic research framework of exploration and exploitation. The Online IM problem still assumes that the diffusion model exists, but the parameters of the model are unknown. On the premise that the parameters of the diffusion model are unknown, we use online learning methods to learn parameters. Interacting with social networks for multiple rounds, so that the parameters can be estimated gradually and accurately, and the optimal seed set can be found to maximize the cumulative influence spread of multiple rounds of interaction. In \cite{lei2015online}, they defined the Online IM problem as: Given a social graph $G=(V,E)$, a diffusion model $\mathcal{D}(G,\vp)$ with unknown $\vp$, a budget $k$, and trial number $T$, it aims to find a set $S_t$ with at most $k$ nodes for each trail $t$ such that $\mathbb{E}[|\cup_{1\leq t\leq T}I(S_t)|]$ is maximized, where $I(S_t)$ is a random variable indicating the influenced node set by seed set $S_t$ in the round $t$. However, this definition is not a standard definition of online problems, and it is impossible to analyze the advantages and disadvantages of the algorithm by theoretical technique. Therefore, we do not consider this definition as the standard definition of the Online IM problem.

\subsection{Problem Definition}\label{sec3-1}
At each round $t\in\{1,2,\cdots,T\}$, we need to select a seed set $S_t$ with at most $k$ nodes from the given social network $G=(V,E)$. Next, starting from $S_t$, spread influence in social networks. We call a round of information diffusion from seed set $S_t$ in the network an influence campaign. After this influence campaign, we will observe the feedback data of this round, and then update the estimation for parameters of the diffusion model according to the feedback. What is feedback data is a debatable question. There are different types of feedback, such as edge feedback and node feedback, which will be discussed in detail in the next section. At the next round $t+1$, we can select a new seed set $S_{t+1}$ based on the diffusion model with updated parameter estimation and repeat the above procedure until round $T$. Therefore, the objective of the Online IM problem is to find a policy to maximize the accumulative influence spread of $T$ rounds. That is
\begin{equation}
\max_{S_1,S_2,\cdots,S_T}\mathbb{E}\left[\sum\nolimits_{t=1}^T|I(S_t)|\right]=\sum\nolimits_{t=1}^T\mathbb{E}\left[|I(S_t)|\right]=\sum\nolimits_{t=1}^T\sigma_{\mathcal{D}}(S_t).
\end{equation}

The research framework of the Online IM problem is based on Multi-Armed Bandit (MAB) in online learning. MAB is a problem extensively studied in statistics and machine learning \cite{berry1985bandit}\cite{sutton2018reinforcement}. 

In the typical setup of a stochastic MAB, there exist $m$ arms, represented as $[m]=\{1,2,\cdots,m\}$. Each arm $i\in[m]$ is associated with a reward distribution that has an unknown mean $\mu_i$. At each round $t$, the agent selects an arm (action) $A_t$, in response to which the environment yields a reward $X_t$ from a distribution that is not known to the agent. The reward received is then used as feedback to inform the agent's future arm selection strategy. The overarching objective for the agent is to maximize the total accumulated reward over time. The MAB framework encapsulates the classic exploration-exploitation trade-off: the decision of whether to continue exploring for potentially better arms or to exploit the best-known arm for rewards. The effectiveness of a policy in this context is typically evaluated by its expected regret, defined as the discrepancy between the expected cumulative reward of consistently choosing the best arm and the rewards obtained under the agent's policy. In this stochastic bandit, the regret of policy $\pi$ around $T$ rounds can be formulated as $R_T(\pi)=T\cdot\max_{i\in[m]}\mu_i-\mathbb{E}[\sum_{t=1}^TX_t]$. Thus, the objective is to minimize the regret and existing results show that $\mathcal{O}(\log T)$ is asymptotically the best possible. For more about bandit algorithms please check Lattimore's monograph \cite{lattimore2020bandit}.

However, the above-mentioned MAB model cannot be directly applied to our online IM problem because we need to select a $k$-size seed set from the social network in each round, where each node set with $k$ nodes will be considered as an arm. In this case, we have $\binom{k}{|V|}$ number of arms, which is hard to realize in polynomial time. Thus, the research of the online IM problem depends on the Combinatorial Multi-Armed Bandit (CMAB) model \cite{audibert2011minimax}\cite{chen2013combinatorial}. In the CMAB model, each arm is called a ``base arm'' and a set of base arms is called a ``super arm''. The agent selects a super arm in each round. According to types of feedback, it can be categorized into full feedback, semi-bandit feedback, and bandit feedback.
\begin{itemize}
    \item \textbf{Full Feedback:} After pulling a super arm, the agent can observe the output of all base arms, regardless of whether the base arm is included in the pulled super arm.
    \item \textbf{Semi-Bandit Feedback:} After pulling a super arm, the agent can only observe the feedback information of base arms contained in the pulled super arm.
    \item \textbf{Bandit Feedback:} After pulling a super arm, the agent can only observe the total reward obtained by pulling this super arm, but cannot see any information of base arms.
\end{itemize}
Here, different feedback types correspond to different application scenarios, in which from full to bandit feedback, the difficulty of solving is gradually increased and the scope of application is correspondingly increased. The Online IM problem is usually modeled by using semi-bandit or bandit feedback. For instance, under edge-level semi-bandit feedback, one can identify which edges are triggered (emanating from nodes that have been influenced). In the case of node-level semi-bandit feedback, it is possible to observe which specific nodes have been influenced. Conversely, with bandit feedback, the observable information is limited to merely the count of nodes that have been influenced.

\subsection{CMAB with Probabilistically Triggered arms (CMAB-T)}
The CMAB model has $m$ base arms, denoted by $[m]$, where each base arm $i\in[m]$ has a distribution of the reward with an unknown mean $\mu_i$. The reward $X_{i,s}$ is a random variable representing the reward of the base arm $i$ at its $s$-th trial, which is independently and identically sampled from a distribution with mean $\mu_i$. At each round $t\in[T]$, the agent selects a super arm $A_t\in\mathcal{S}\subseteq 2^{[m]}$, which will trigger all base arms included in $A_t$. In addition, considering the CMAB-T model \cite{chen2016combinatorial}, the triggered base arms in $A_t$ may trigger other base arms not in $A_t$, and these triggered arms may further trigger more base arms, and so on. Let $p^i_{A_t}$ be the triggering probability of base arm $i$ if the super arm $A_t$ is selected and $T_{i,t}$ be the number of times arm $i$ has been triggered in the first $t$ rounds. Thus, the reward in the round $t$, denoted by a random variable $R_t(A_t)$, can be a (linear or non-linear) function of rewards $X_{i,T_{i,t}}$ for each triggered base arm $i$ in this round. Taking linear reward as an example, we have $R_t(A_t)=\sum_{i\text{ is triggered}}X_{i,T_{i,t}}$. When an arm $i$ is triggered at round $t$, we can use the observed reward $X_{i,T_{i,t}}$ to update its mean estimation as $\hat{\mu}_i=\sum_{s=1}^{T_{i,t}}X_{i,s}/T_{i,t}$. The super arm that is expected to give the highest reward is selected in each round by an oracle $O$. The oracle $O$ takes the current mean estimation $\hat{\boldsymbol{\mu}}=(\hat{\mu}_1,\hat{\mu}_2,\cdots,\hat{\mu}_m)$ as input and return an super arm. In \cite{chen2013combinatorial}\cite{chen2016combinatorial}, they assume a $(\alpha,\beta)$ oracle that returns an $\alpha$-approximation solution with at least $\beta$ probability.

\begin{table}[!t]
		\renewcommand{\arraystretch}{1.2}
		\label{table1}
		\centering
        \resizebox{\linewidth}{!}{
		\begin{tabular}{l|c|l}
			\hline
			CMAB & Symbol & Mapping to the online IM\\
			\hline
		  Base arm & $i$ & Edge $(u,v)$\\
            Reward for arm $i$ in its $s$-th trail & $X_{i,s}$ & Activation status (live/dead) for edge $(u,v)$\\
            Mean of distribution for arm $i$ & $\mu_i$ & Diffusion probability $p_{uv}$\\
            Super arm at round $t$ & $A_t$ & Union of outgoing edges from seed nodes in $S_t$\\
            Triggered times for arm $i$ until round $t$ & $T_{i,t}$ & \# of times $u$ is activated in the first $t$ influence campaign\\
            Reward in round $t$ & $R_t$ & Influence spread $|I(S_t)|$ in the $t$-th influence campaign\\
			\hline
		\end{tabular}}
        \caption{The mapping from the CMAB model to the Online IM problem.}
	\end{table}

Back to the Online IM problem under the IC model, it can be modeled by using CMAB-T naturally. The mapping from CMAB to the Online IM problem is shown in Table \ref{table1}. In each round $t$, the agent selects a seed set $S_t$ with $|S_t|\leq k$, which implies playing the corresponding super arm $A_t$ composing of all outgoing edges from nodes in this seed set $S_t$. Here, $S_t$ can be selected by randomly exploring or by exploiting current diffusion probability estimation $\hat{\boldsymbol{\mu}}$ to solve the Offline IM problem. Given a social graph $G$ and diffusion model $\mathcal{D}(\hat{\boldsymbol{\mu}})$, the Offline IM algorithms as shown in Section \ref{sec2-3}, such as IMM \cite{tang2015influence} and OPIM-C \cite{tang2018online}, can work as a $(1-1/e-\varepsilon,1-1/|V|)$ oracle to solve the Offline IM problem. When the super arm $A_t$ is played ($S_t$ is selected), information begins to diffuse across the networks, and then the agent can observe a realization $g_t$. The activation statuses of edges in $E_{g_t}$ can be observed. Since each edge in $E_{g_t}$ is either live or dead in the realization, we can assume a Bernoulli distribution $X_{i,s}\in\{0,1\}$. Thus, in this round, $X_{i,T_{i,t}}=1$ if arm $i$ (edge $(u,v)$) is live; $X_{i,T_{i,t}}=0$ if arm $i$ (edge $(u,v)$) is dead. The final reward of round $t$ is the number of active nodes at the end of the diffusion process, and thus we have $R_t(A_t)=|I(S_t)|$ which is a non-linear function of rewards for the triggered arms (observed edges).

After an influence campaign, the diffusion probability estimation needs to be updated. For the realization $g_t$ in the round $t$, each arm (edge) in $E_{g_t}$ has been triggered. Thus, for each arm $i\in E_{g_t}$, we have $\hat{\mu}_i=\sum_{s=1}^{T_{i,t}}X_{i,s}/T_{i,t}$. Since the Offline IM problem is NP-hard problem, even if the diffusion probability $\vp$ ($\boldsymbol{\mu}=(\mu_1,\mu_2,\cdots,\mu_m)$) is known, as shown in Section \ref{sec2-3}, the best expected influence spread we can obtain is $\alpha\cdot\beta\cdot\max_{|S|\leq k}\sigma_{\mathcal{D}(\vp)}(S)$ where $\alpha=1-1/e-\varepsilon$ and $\beta=1-1/|V|$. Thus, the expected regret of policy $\pi$ can be written as
\begin{equation}
    R_T(\pi)=T\cdot\alpha\cdot\beta\cdot\max_{|S|\leq k}\sigma_{\mathcal{D}(\vp)}(S)-\mathbb{E}\left[\sum\nolimits_{t=1}^T|I(S_t)|\right].
\end{equation}

In summary, Chen et al. \cite{chen2013combinatorial}\cite{chen2016combinatorial} is the first to study the Online IM problem by using the CMAB model, which is an instance of semi-bandit feedback based on edge-level feedback. They gave a complete theoretical analysis of expected regret. This is a pioneering work about the Online IM problem, and later, some research work on this subject is also based on this model.

\section{Summarization of Online IM Algorithms}\label{sec4}
In this last section, we have seen that the Online IM problem is an instance of the CMAB-T model. According to the taxonomy of the CMAB model mentioned in Section \ref{sec3-1}, it belongs to semi-bandit feedback based on edge-level feedback. In fact, such edge-level feedback is an idealized situation, where we need to accurately observe the interaction between each pair of nodes. This condition is undoubtedly harsh, and subsequent studies have noticed other types of feedback that are more practical. Therefore, in this section, we will expand to summarize Online IM algorithms according to different types of feedback, divided into edge-level, node-level, and other weaker feedback.

\IncMargin{1.5em}
\begin{algorithm}[!t] 
    \caption{Combinatorial Upper Confidence Bound (CUCB) Algorithm}
    \label{a1}
    Initialize: $T_i=0$, $\hat{\mu}_i=1$ for each $i\in[m]$\;
    \For{$t=1$ to $T$}{
        $\bar{\mu}_i\leftarrow\min\{\hat{\mu}_i+\sqrt{{3\ln t}/{2T_i}},1\}$ for each $i\in[m]$\;
        $S_t\leftarrow\text{Oracle}(\bar{\mu}_1,\bar{\mu}_2,\cdots,\bar{\mu}_m)$; //Offline IM algorithms\\
        Play $S_t$ ($A_t$), observe outcomes of triggered based arms $E_t\subseteq E$, and $T_i\leftarrow T_i+1$ for each $i\in E_t$\;
        Update $\hat{\mu}_i\leftarrow(\hat{\mu}_i\cdot(T_i-1)+X_{i,T_i})/T_i$ for each $i\in E_t$\;
    }
\end{algorithm}

\subsection{Edge-level Feedback Models}
Based on the CMAB-T model, Chen et al. \cite{chen2013combinatorial}\cite{chen2016combinatorial} proposed a Combinatorial Upper Confidence Bound (CUCB) algorithm as Shown in Algorithm \ref{a1}. The CUCB algorithm is very similar to the standard UCB algorithm in the MAB model \cite{auer2002finite}. It maintains a trigger times $T_i$ and empirical mean $\hat{\mu}_i$ for each base arm $i\in[m]$. In each round $t\in[T]$, the agent regards the confidence upper bound $\bar{\mu}_i$ for each arm $i\in[m]$ as the diffusion probability and inputs them to the offline oracle, namely Offline IM algorithms, to get a suboptimal seed set $S_t$. Here, we have $\sigma_{\mathcal{D}(\bar{\boldsymbol{\mu}})}(S_t)\geq\alpha\cdot\max_{|S|\leq k}\sigma_{\mathcal{D}(\bar{\boldsymbol{\mu}})}(S)$ with at least $\beta$ probability. After selecting the seed set $S_t$ (executing the action $A_t$), the agent can update the empirical mean and confidence upper bound for each arm, which will be used to select a new seed set $S_{t+1}$ in the next round. The research indicated that the distribution-dependent regret for the CUCB algorithm is constrained within $\mathcal{O}(\log T)$, and they also extended this to offer distribution-independent regret bounds for a wide array of reward functions. However, a notable limitation of the CUCB's regret bound is the inclusion of a factor $1/p^*$, which can be exponentially large in certain cases, where $p^*$ is defined as the smallest positive probability of triggering a base arm through any chosen action. This significant drawback was addressed in \cite{wang2017improving} by integrating a Triggering Probability Modulated (TPM) bounded smoothness condition into the broader CMAB-T model. It was also demonstrated that the Online IM problem adheres to this TPM condition. Thus, the regret bound for the Online IM problem based on the CMAB-T model can be $\Tilde{\mathcal{O}}(|V||E|\sqrt{T})$, where the $\mathcal{\Tilde{O}}$ notation is often used to ignore logarithmic factors, which are significantly improved than before.

The CMAB-T model and its corresponding CUCB algorithm completely follow the basic assumption given by the IC model. That is to say, each edge has a fixed diffusion probability and is independent of each other. Subsequently, some researchers made further expansions based on this. Wen et al. \cite{wen2017online} made a linear generalization to the diffusion probability in the IC model because they considered that the number of edges in real social networks is too large to develop efficient online learning algorithms. Thus, they assume that it exists an unknown global parameter $\boldsymbol{\theta}^*\in\mathbb{R}^d$ and a known local parameter $\boldsymbol{x}_e\in\mathbb{R}^d$ for each edge $e\in E$, and the diffusion probability $p_e$ of each edge $e\in E$ can be well approximated by $\boldsymbol{x}_e\cdot\boldsymbol{\theta}^*$, where $\boldsymbol{x}_e$ is the feature vector of edge $e$ determined by its intrinsic attribute. Thus, the problem is transformed into an estimation of the global vector $\boldsymbol{\theta}^*$, which is greatly simplified. Based on this linear bandit, they proposed an IMLinUCB algorithm by using a process similar to the CUCB algorithm. The IMLinUCB obtains a faster learning rate to estimate the global parameter with time complexity $\mathcal{O}(|E|d^2)$ to update parameters and reaches a similar upper bound $\Tilde{\mathcal{O}}(d|V||E|\sqrt{T})$ of cumulative regret with the CUCB. Later, based on the edge-level feedback in CUCB and linear generalization in IMLinUCB, Vaswani et al. \cite{vaswani2017model} re-parameterized the Online IM problem in a model-independent manner. The approach involved refining a surrogate objective $f(S,\vp^*)=\sum_{v\in V}f(S,v,\vp^*)$, which is based on the concept of pairwise reachability. Here, the $f(S,v,\vp^*)=\max_{u\in S}p_{uv}^*$ is defined as the maximum pairwise reachability from the seed set $S$ to a specific node $v$. Here, the $p_{uv}^*$ is the pairwise reachability from node $u$ to $v$, which is the likelihood of $v$ being influenced assuming $u$ is the sole seed node in the graph $G$ under the diffusion model $\mathcal{D}$. Now, the pairwise reachability for each edge $(u,v)\in E$ can be approximated by $\boldsymbol{\theta}^*_u\cdot\vx_v$, where $\boldsymbol{\theta}^*_u\in\mathbb{R}^d$ is the (unknown) source weight and $\vx_v\in\mathbb{R}^d$ is the (known) target feature. Based on this linear bandit, they proposed a DILinUCB (Diffusion-Independent LinUCB) algorithm by using a process similar to the CUCB algorithm to estimate the source weight parameters. The regret has also been improved in this work. Even though they claimed that it can adapt to different diffusion models, the linear reachability assumption may violate the diffusion rules in some diffusion models, such as the IC model. Besides, the objective of the DILinUCB is only an approximation to the real objective of the Online IM problem, which is equivalent to changing the definition of the original problem. The cost of this overly heuristic method is sometimes unacceptable, which has no theoretical guarantee about its approximation quality.

Following this line of thought, Wu et al. \cite{wu2019factorization} made further explorations by considering network assortativity. The diffusion probability $p_{uv}$ of each edge $(u,v)\in E$ can be factorized as $p_{uv}=\boldsymbol{\theta}_u\cdot\boldsymbol{\beta}_v$, where $\boldsymbol{\theta}_u\in\mathbb{R}^d$ is the influence factor associated with node $u$ and $\boldsymbol{\beta}_v\in\mathbb{R}^d$ signifies the susceptibility factor of node $v$. Now, the model complexity should be $\mathcal{O}(d|V|)$. Based on this linear bandit, they proposed an IMFB algorithm by using a process similar to the CUCB algorithm to estimate the influence and susceptibility parameters. Compared with the model complexity $\mathcal{O}(|E|)$ in CUCB and IMLinUCB, its model complexity $\mathcal{O}(d|V|)$ has been significantly reduced. Even though the model complexity of DILinUCB is also $\mathcal{O}(d|V|)$, it is only a heuristic as we said before. At the same time, the upper regret bound of IMFB can be improved to $\mathcal{O}(d|V|^{5/2}\sqrt{T})$ compared with $\mathcal{O}(d|V|^{3}\sqrt{T})$ in IMLinUCB and $\mathcal{O}(|V|^{3}\sqrt{T})$ in CUCB respectively. However, all these works \cite{wen2017online}\cite{vaswani2017model}\cite{wu2019factorization} predefined a linear assumption on parameters for estimating edge probabilities, where The probability matrix, as estimated, represents a low-rank approximation of the actual parameter matrix, thereby confining the model within a constrained low-rank space. Such an approach is subject to significant limitations.

\subsection{Node-level Feedback Models}
Even though the bandit problem with edge-level feedback is straightforward to get a theoretical guarantee, the update according to the information from edge-level feedback is not realistic in many practical applications since we need to observe the statuses of all connections between users, which is too difficult. Thus, there are some follow-up researchers attempting to overcome the drawback of edge-level feedback by introducing more practical node-level feedback. The node-level feedback only discerns whether each node has been activated, rather than identifying which specific node was responsible for this activation.

Based on the CMAB-T model, Vaswani et al. \cite{vaswani2015influence} was the first to consider the Online IM problem based on node-level feedback, where updating the mean estimate for each edge presents a greater challenge due to the uncertainty about which active predecessor was responsible for the activation of a node. Any of the active parents may activate a node, which leads to a credit assignment problem. They proposed two different ways to resolve this problem by transforming node-level feedback to edge-level feedback. The first one is using Maximum Likelihood Estimation (MLE) to build a log-likelihood function for a given set of cascades $C$. The second one is the Frequentist approach, which chooses one of the active parents randomly and assigns the credit of this activation to it since the diffusion probabilities are usually small in typical social networks. However, this work fails to give the performance analysis of the Online IM problem under the node-level feedback model.

With the great success of the CMAB model in studying the Online IM problem under the IC model, there were some researchers beginning to consider other diffusion models. Li et al. \cite{li2020online} was the first to study the Online IM problem under the LT model, which is based on the node-level feedback since node activation under the LT model is aggregated from all active incoming neighbors. Here, the (full) node-level feedback refers to the agent can observe the influence diffusion process, activated node sets in all timestamps denoted by a diffusion sequence $\{S_{t,0},\cdots,S_{t,\tau},\cdots,S_{t,|V|-1}\}$ at round $t$, which is different from the default node-level feedback, knowing only the set of nodes activated by the end of the diffusion process. Motivated by linear bandits used in the edge-level feedback \cite{wen2017online}\cite{vaswani2017model}\cite{wu2019factorization}, they proposed a LT-LinUCB algorithm updated by using the distilled information. When satisfying the Group Observation Modulated (GOM) bounded smoothness, similar to the TPM bounded smoothness \cite{wang2017improving} under the IC model with edge-level feedback, the regret can be bounded within $\mathcal{O}(poly(|E|)\sqrt{T}\log T)$. Besides, they gave an Explore-Then-Commit (ETC) algorithm, OIM-ETC, which can be applied to both IC and LT models with node-level feedback. According to this important finding, there is a nearly optimal $\mathcal{\Tilde{O}}(\sqrt{T})$-regret algorithm for the Online IM problem with (full) node-level feedback. Zhang et al. \cite{zhang2022online} continued to study whether it exists a similar conclusion under the IC model at the cost of involving offline pair oracles (full node-level feedback) instead of standard oracles (partial node-level feedback). By using a similar process to Vaswani's work \cite{vaswani2015influence}, they extracted information to estimate diffusion probability $p_{uv}$ for each $(u,v)\in E$ and its confidence ellipsoid from the feedback $\{S_{t,0},\cdots,S_{t,\tau},\cdots,S_{t,|V|-1}\}$ at round $t$ by using a novel adaptation of the MLE approach. Then, they proved the GOM bounded smoothness under the IC model and proposed a $\mathcal{\Tilde{O}}(\sqrt{T})$-regret algorithm under the IC model with offline pair oracles. However, for both LT \cite{zhang2022online} and IC \cite{zhang2022online} models, they did not get a nearly optimal Online IM algorithm under standard offline oracles, and there is no general bandit framework like CMAB-T that can take the Online IM problem as a special case.

In situations where multiple edges simultaneously attempt to activate the same subsequent node, neither UCB-based nor Thompson Sampling (TS)-based algorithms can find an accurate estimation of every single edge, just the aggregated probabilities. Yang et al. \cite{yang2019online}\cite{yang2021online} considered the Online IM problem under (partial) node feedback, where they assumed the diffusion probability is $p_e=1-\exp(-\vx_e^T\boldsymbol{\theta}^*)$ for each edge $e\in E$. Like this, the likelihood function is concave so that easy to be maximized. They are the first to construct a confidence region of the MLE and use this confidence region to propose an online learning algorithm with $\mathcal{\Tilde{O}}(\sqrt{T})$ cumulative regret, which matches the corresponding result in edge-level feedback \cite{wen2017online}.

\subsection{Other Feedback Models}
In the previous two subsections, we introduce edge-level semi-bandit and node-level semi-bandit IM algorithms respectively. In this subsection, we are mainly concerned about other weaker feedback models, such as bandit feedback. Bao et al. \cite{bao2016online} made two innovations here. First, they focused on a dynamic non-stationary social network, where diffusion probabilities and network topology are unknown and varying over time. Thus, the weak regret can be defined as
\begin{equation}
    R_T(\pi)=T\cdot(1-1/e)\cdot\max_{|S|\leq k}\sum\nolimits_{t=1}^T\mathbb{E}\left[f_t(S)\right]-\sum\nolimits_{t=1}^T\mathbb{E}\left[f_t(S_t)\right],
\end{equation}
where $f_t(S)$ is an influence spread function in round $t$. Second, they assumed that the agent could only observe the number of activated nodes after each influence campaign. They defined $a|S$, node $a$ under a given set $S$, as an arm and used a greedy algorithm to gradually add the node with a maximum marginally influence spread. They proposed an RSB (Randomized Sequential MAB algorithm for non-stationary networks) algorithm. It followed the Exp3 algorithm in the adversarial MAB model which maintains a component of exploitation and exploration for each node and selects one by one to join the seed set.

Besides, there are two works considering the Online IM problem with local observation \cite{carpentier2016revealing}\cite{lugosi2019online}. Carpentier et al. \cite{carpentier2016revealing} considered a local influence structure in an unknown underlying graph, where each node can only influence its neighbors. They proposed a UCB-based BARE algorithm and its performance guarantee did not scale with the number of nodes. However, this approach is tantamount to simplifying the problem, and whether the result can maximize the overall influence spread is debatable. Lugosi et al. \cite{lugosi2019online} assumed that the agent can only observe the out-degree of the seed node after selecting a seed set, and they asserted that these localized observations are adequate for optimizing overall influence within two widely analyzed families of random graph models: stochastic block models and Chung–Lu models. Thus, they proposed a UCB-based algorithm to maximize the local influence spread and then maximize the global influence spread. Furthermore, Bayiz et al. \cite{bayiz2022decentralized} thought that it is unrealistic for a central observer can globally monitor the influence spread. Thus, they introduced two decentralized Online IM algorithms, ETC and $\varepsilon$-greedy algorithms, that delegate the online updates to the nodes on the network, which requires only local observations at the nodes. They proved that the proposed ETC algorithm has a $\mathcal{O}(T^{2/3})$ cumulative regret when diffusion probabilities for each connection in the network are symmetric. It indicates that this theoretical guarantee can only be applied to undirected graphs.

Recently, Agarwal et al. \cite{agarwal2022stochastic} further consider the Online IM problem where only the amount of influence is needed (No other feedback like activated nodes or edges is available.) Here, this only feedback is a nonlinear reward of the selected $K$ arms. For an action $\va_t=(a_{1,t},a_{2,t},\cdots,a_{K,t})$, let $d_{\va_t}=(X_{a_1,t},X_{a_2,t},\cdots,X_{a_K,t})\in[0,1]^K$ be arm rewards at round $t$ from action $\va_t$. Thus, the reward is $r_{\va}(t)=f(d_{\va_t})$ and the expected regret is $T\cdot r_{\va^*}-\mathbb{E}[\sum_{t=1}^Tr_{\va_t}(t)]$. According to this setting, they proposed a CMAB-SM algorithm based on the divide-and-conquer strategy and achieved a $\mathcal{\Tilde{O}}(K^{1/2}|V|^{1/3}T^{2/3})$ regret bound. In their simulations, it showed that CMAB-SM gave an excellent performance in the Online IM problem, which is better than similar algorithms with the same degree of feedback in \cite{goyal2011data}\cite{agarwal2021dart}. In my opinion, this kind of feedback, namely we only know the number of influenced nodes at the end of the diffusion process, maybe the most practical. For example in advertisement placing, a company or algorithm designer is easy to know how many consumers eventually buy this product. Edge-level or node-level diffusion information may not be made public for the purpose of privacy protection, and even if it is made public, the cost of obtaining such standardized diffusion data is extremely high. However, the research based on such bandit feedback is very limited, thus this is the focus of future research in this area.

\section{Further Studies of Online Influence Maximization}\label{sec5}
Due to the complexity and applicability of the Online IM problem, a lot of researchers continued to study this field deeply. One direction is to use more sophisticated models and mathematical theories to consider the Online IM problem from another perspective. The other direction is to study variants of the Online IM problem motivated by more realistic applications, such as Topic-Aware and Location-Aware IM problems. Therefore, in this section, we start from the following two aspects.

\subsection{Innovative Modeling}
Looking back on the seminal work \cite{lei2015online} of the Online IM problem, we mentioned in Section \ref{sec3} that its definition is not canonical from the perspective of online learning because the reward of a round is the number of newly activated nodes. Actually, this is an online version of the Multi-Round IM problem \cite{sun2018multi}. The Multi-Round IM problem is offline, but it appeared even later, Sun et al. \cite{sun2018multi} formulated this problem and gave two efficient approximation algorithms to solve it in 2018. Based on the incremental definition in \cite{lei2015online}\cite{sun2018multi}, Lagr{\'e}e et al. \cite{lagree2018algorithms} considered to maximize the $\mathbb{E}[\cup_{1\leq t\leq T}I(S_t)]$ by parameterize nodes, which represent the potentials of remaining spreads from each of the influencer nodes. They proposed a GT-UCB algorithm based on the Good-Turing (GT) estimation to sequentially select the seeds to activate at each round, whose performance is competitive with other Online IM algorithms. Based on a similar objective function, by assuming contextual information is known and exploitable in the learning process, Iacob et al. \cite{iacob2022contextual} proposed two diffusion-model agnostic UCB-based algorithms, GLM-GT-UCB and LogNorm-LinUCB, by using Poisson and log-normal assumptions on the underlying distributions of the number of influenced nodes. The GLM-GT-UCB uses a Generalized Linear Model (GLM) and GT estimator for new activation and the LogNorm-LinUCB adapts the LinUCB to this setting directly. They exhibit different and complementary behaviors according to different application scenarios in real-world data simulations.

By adopting the Thompson Sampling (TS) technique, Wang et al. \cite{wang2018thompson} first studied the applications of the TS to the CMAB model and proposed a Combinatorial Thompson Sampling (CTS) algorithm that is standard TS policy with the offline oracle. Then, H{\"u}y{\"u}k et al. \cite{huyuk2019analysis}\cite{huyuk2020thompson} integrated the CTS algorithm into the CMAB-T model \cite{chen2016combinatorial}, where the agent maintains a posterior distribution over the expected rewards of each arm and selects a super arm based on the posterior of each arm by utilizing an offline oracle at each round. When the expected reward adheres to the TPM Lipschitz continuity, the CTS algorithm attains a Bayesian regret of $\mathcal{O}(\max\{m\sqrt{T\log T},m^2\})$. Furthermore, in scenarios where all base arms have non-zero triggering probabilities, CTS achieves a regret of $\mathcal{O}(1/p^*\log(1/p^*))$, independent of the time horizon, with $p^*$ being the smallest non-zero triggering probability. A comparison with the CUCB algorithm in the Online IM problem with edge-level semi-bandit feedback reveals similar outcomes when both algorithms employ an approximation oracle, as opposed to an exact oracle.

In previously mentioned linear bandits \cite{wen2017online}\cite{vaswani2017model}\cite{wu2019factorization}, diffusion probabilities are fitted based on independent edge/node's features, ignoring the correlation with other nodes/edges on the graph. Thus, Xia et al. \cite{xia2020gaussian} estimated diffusion probabilities based on the Gaussian Process (GP) model, where it needs to learn a general function such that $p_{uv}=g(\vx_{uv})$ for each $(u,v)\in E$ and $\vx_{uv}^T=[\boldsymbol{\theta}_u^T,\boldsymbol{\theta}_v^T]$ like \cite{wu2019factorization}. In this GP model, the estimation $g(\vx_{uv})$ is a random variable following a posterior Gaussian distribution. Based on this model, they proposed an IMGUCB (IM using Gaussian Process and UCB) algorithm with a similar process as before. They gave a theoretical analysis of expected regret and validated that it circumvents the drawbacks of previous linear models and generalizes to arbitrary diffusion probability patterns in the experiment. Besides, However, both CUCB \cite{chen2016combinatorial} and CTS \cite{huyuk2020thompson} do not exploit correlations between base arms during the learning process. Demirel et al. \cite{demirel2021combinatorial} consider the CMAB-T model where the expected reward of the base arm is a sample from the GP and the offline oracle is exact. GPs can model the dependencies between the expected rewards of different base arms by using a kernel function. They showed that the ComGP-UCB (Gaussian Process Upper Confidence Bound) algorithm proposed in \cite{srinivas2010gaussian} achieves $\mathcal{O}(m\sqrt{T\log T/p^*})$ regret bound with a high probability under the TPM Lipschitz continuity on the expected reward. They claimed that the ComGP-UCB is better than CUCB and CTS when the expected rewards of base arms are not independent. Even though that, the $1/p^*$ term in regret cannot be removed, and this algorithm was not tested in a real Online IM problem.

\subsection{Variants in Real Applications}\label{sec5-2}
Motivated by the Topic-Asware IM problem \cite{chen2015online}, Sar{\i}ta{\c{c}} et al. \cite{saritacc2016online} considered an Online Contextual IM Problem (OCIMP), where there is a context (topic) $\vx_t\in\mathbb{R}^d$ in each round $t$, which can determine the diffusion probability $\vp^{\vx_t}$ in this round. Thus, the expected regret would become $R_T(\pi)=\alpha\beta\cdot\sum_{t=1}^T\sigma(\vx_t,S^*(\vx_t))-\sum_{t=1}^T\sigma(\vx_t,S_t)$, where $S^*(\vx)\in\arg\max_{|S|\leq k}\sigma(\vx,S)$ is the offline oracle give the context $\vx$. Building upon this contextual bandit, they introduced a COIN algorithm, which achieves a sublinear regret bound when the influence probabilities are Hölder continuous with respect to the context. Recognizing a gap in previous research, which overlooked the cost associated with feedback (specifically, the cost of observing the influence spread), they \cite{saritacc2018online} expanded their work to include the Online Contextual IM Problem with Costly Observations (OCIMP-CO). This extension considers two distinct observation scenarios: costly edge-level feedback (where observing influenced nodes is free, but there is a cost for observing activation on edges) and costly node-level feedback (where there is a cost to determine if a node has been influenced). Thus, the regret would become $R_T(\pi)=\alpha\beta\cdot\sum_{t=1}^T\sigma(\vx_t,S^*(\vx_t))-\sum_{t=1}^T\sigma(\vx_t,S_t)+c\sum_{t=1}^TB_t$ where $B_t$ is the number of observations at round $t$. For the costly edge-level feedback, they proposed a COIN-CO-EL algorithm with $\mathcal{O}(c^{1/3}T^{(2\theta+d)/(3\theta+d)})$ regret when influence probabilities are Hölder continuous functions of the context, where $c$ is the observation cost and $\theta$ is the exponent of Hölder condition. For the costly node-level feedback, they proposed a COIN-CO-NL algorithm with $\mathcal{O}(c^{1/3}T^{(2\theta+d)/(3\theta+d)})$. When the diffusion probabilities are independent of the context, the regret can be improved to $\mathcal{O}(c^{1/3}T^{2/3})$.

Motivated by the Budgeted IM problem \cite{nguyen2013budgeted}\cite{bian2020efficient}, some researchers began to talk about online versions of the Budget IM problem under edge-level semi-bandit feedback. Wang et al. \cite{wang2020online} first proposed a cumulative oversampling method that can be applied to the Online IM problem, which is inspired by the oversampling idea for Thompson Sampling \cite{russo2018tutorial}. Then, they applied the cumulative oversampling to the Online Budget IM problem called Lin-IMB-L (Budgeted IM Semi-Bandit with Linear Generalization of Edge Weights), where it was predicated on a global budget constraint, formulated as $\mathbb{E}[\sum_{t=1}^Tc(S_t)]\leq B$ and $c(\cdot)$ is a cost function. They integrated the cumulative oversampling method with offline approximation oracles from the Budgeted IM problem into an online learning framework. It was demonstrated that the regret bound of this approach is $\mathcal{\Tilde{O}}(d|V||E|\sqrt{T})$, aligning with the regret bound of the UCB-based algorithm in the same parameterized context as seen in IMLinUCB \cite{wen2017online}. Based on the above-mentioned global budget, Perrault et al. \cite{perrault2020budgeted} formally articulated the Online Budgeted IM problem, assuming an IC model and edge-level semi-bandit feedback. They proposed a performance metric for online policies akin to the approximation regret in \cite{chen2013combinatorial}. A UCB-based algorithm was developed, showing that its regret could be logarithmically bounded. This online policy is also adaptable for solving the Online IM problem, offering an enhancement over the conventional analysis of the CUCB algorithm \cite{wang2017improving}.

\begin{table}[!t]
		\renewcommand{\arraystretch}{1.2}
		\label{table2}
		\centering
        \resizebox{\linewidth}{!}{
		\begin{tabular}{l|l|l|l|l|l}
			\hline
			Algorithm & Prior? & Offline computation & Feedback & Type & Regret\\
			\hline
                OCIM-TC & $\checkmark$ & Standard & Full propagation & Bayesian & $\mathcal{O}(\sqrt{T\ln T})$\\
                OCIM-OFU & $\times$ & Hard & Full propagation & Frequentist & $\mathcal{O}(\sqrt{T\ln T})$\\
                OCIM-ETC & $\times$ & Standard & Direct out-edges & Frequentist & $\mathcal{O}(T^{2/3}(\ln T)^{1/3})$\\
			\hline
		\end{tabular}}
        \caption{Summary of proposed algorithms for the Online CIM problem \cite{zuo2022online}.}
	\end{table}

The Competitive IM problem \cite{bharathi2007competitive}\cite{guo2020multi} is one of the important variants in the family of the IM problem, where two competing items diffuse in the same network. Zuo et al. \cite{zuo2022online} was the first to consider the Competitive IM problem in an online setting based on the Competitive IC model \cite{bharathi2007competitive} where diffusion probabilities for both items are unknown. They introduced a general Contextual Combinatorial Multi-Armed Bandit with Probabilistically Triggered Arms ($\text{C}^2$MAB-T) for the Online Competitive IM problem, where the monotonicity property (influence spread increases when influence probabilities on edges increase) mentioned in \cite{chen2016combinatorial} no longer holds, thus the UCB-based algorithm cannot be used, but the TPM bounded smoothness condition in CMAB-T holds, which points out the design direction of the algorithm. Thus, they provided three solutions with different tradeoffs between prior knowledge, offline computation, feedback, and regret bound: OCIM-TS, OCIM-OFU, and OCIM-ETC, as shown in Table \ref{table2}. Here, the regret analysis of OCIM-TS combines the TS with the TPM bounded condition to overcome the difficulty of non-monotonicity and allows any oracles; and OCIM-OFU and OCIM-ETC extend the analysis from CMAT to $\text{C}^2$MAB-T.

Based on the Combinatorial Contextual Bandit ($\text{C}^2$MAB) model \cite{li2016contextual}\cite{wen2017online} and the $\text{C}^2$MAB-T model \cite{zuo2022online} designed for the Online Competitive IM problem, Liu et al. \cite{liu2023contextual} formally defined a $\text{C}^2$MAB-T model, where a time-dependent feature map, denoted as $\phi_t$, is employed to encapsulate the contextual information for each round $t$. The mean reward for any given arm $i\in[m]$ is determined by a linear product of $\phi_t(i)\in\mathbb{R}^d$ and $\boldsymbol{\theta}^*\in\mathbb{R}^d$. This is different from the $\text{C}^2$MAB-T model \cite{zuo2022online} in the Online Competitive IM problem because their meaning of ``contexts'' is the action of the competitor. They designed a $\text{C}^2$-UCB-T algorithm that can be proved $\Tilde{\mathcal{O}}(d\sqrt{KT})$ when satisfying the TPM bounded smoothness condition. Then, they proposed a VA$\text{C}^2$-UCB algorithm that can be proved $\Tilde{\mathcal{O}}(d\sqrt{T/p_{min}})$ when satisfying the Variance Modulated (VM) smoothness condition \cite{liu2022batch}. Finally, they further achieved $\Tilde{\mathcal{O}}(d\sqrt{T/})$ regret by investigating the triggering probability and Variance Modulated (TPVM) condition \cite{liu2022batch}, where they found that the proposed Triggering Probability Equivalence (TPE) analysis and variance-adaptive algorithm can be used to improve existing results on CMAB-T as well.

However, previous research about the Online IM problem was only based on IC or LT model, they failed to consider the common decaying phenomenon in influence diffusion \cite{kempe2005influential}\cite{zhang2016influence}. Thus, Fang et al. \cite{10.5555/3545946.3598895} formulated the Online IM problem under the Decreasing Cascade (DC) model, where the activation probability of $u\in N^-(v)$ on $v$ decreases with more previous failed attempts and the probability that $v$ is influenced does not depend on the order of these nodes' activation attempts. They proposed a UCB-based algorithm, DC-UCB, that can not only maintain the decreasing property of DC but also achieve the same regret bound under the IC model by proving a DC-based TPM bounded smoothness condition. 


\section{Challenges and Future Research Directions}\label{sec6}
The Online IM problem is a special kind of CMAB problem, thus it inherits the difficulties and challenges of the CMAB problem shown as follows: (1) \textbf{Complex Arm Structures:} In CMAB, arms can be viewed as forming combinatorial structures, such as bipartite graphs. This complexity means that arms are not played in isolation but as part of a set, which significantly increases the complexity of the problem. (2) \textbf{Non-Linear Reward Structures:} The reward structure in CMAB is not a simple linear function of the outcomes of all played arms but can take a more complicated form. This non-linearity adds another layer of complexity to the problem. (3) \textbf{Exponential Number of Super Arms:} Treating every super arm as an arm in classical MAB frameworks can lead to combinatorial explosion, with the number of super arms growing exponentially with the size of the problem instance. This makes classical MAB algorithms less effective and efficient. (4) \textbf{Probabilistically Triggered Arms:} The CMAB framework has been extended to include arms that are probabilistically triggered, further complicating the learning and optimization process. Therefore, for the CMAB problem, not only does it need to focus on designing enhanced algorithms to better handle the intricacies of the CMAB problem, but also defines a more general framework that can encompass a wide range of nonlinear reward functions and accommodate various complexities.

As far as the current research development is concerned, the Online IM problem is based on the basic framework laid by CMAB. However, due to its inherent complexity, such as the spread of influence, there are still many problems that need further study in the future. Therefore, in this section, we mainly summarize them into the following four aspects.
\begin{itemize}
    \item While its parameters are unknown, the diffusion model must be assumed in advance. For example, the current mainstream CMAB model \cite{chen2016combinatorial}\cite{wang2017improving} is based on edge-level semi-bandit feedback, whose potential assumption is the diffusion process conforms to the independence hypothesis defined by IC model. However, in real-world social applications, the diffusion mechanism is complex and diverse. At present, the parameter estimation methods are not only based on the independence hypothesis but also on the linear reward hypothesis, such as a series of linear bandit methods. This has great limitations. On the other hand, node-level semi-bandit feedback and bandit feedback, which are more in line with real applications, are more challenging to update the diffusion model, because we cannot know which neighbor was successfully activated at which time in the propagation process. Thus, in current research, algorithms based on node-level feedback usually need to know the activated nodes at each timestamp during the diffusion process, which undoubtedly greatly increases the difficulty of sampling and limits its application range. For the bandit feedback, there is no effective means to obtain a theoretical guarantee until now. Therefore, it is a feasible research direction in the future to make full use of the feedback information that is more in line with the actual situation, make full use of the limited samples, and increase the efficiency of the samples.
    \item The Online IM problem is very different from the Offline IM problem because the online problem is more sensitive to the given diffusion model. In the Offline IM problem, different diffusion models can satisfy that the influence functions defined by them have the same properties, such as monotonicity and submodularity, and their optimization algorithms are compatible with each other. For example, IMM \cite{tang2015influence} was designed to solve the IM problem under the IC model, but it can adapt to the LT model with a minor modification. However, in the nline IM problem, such convenience will no longer exist. Online IM algorithms designed for the IC model cannot applied to other diffusion models since the essence of online algorithms is a parameter estimation problem, but the parameter form and distribution of each diffusion model are totally different. It indicates that the online algorithm must be designed according to a given specific diffusion model. The parameter estimation of IC model is the simplest one, and how to solve the Online IM problem on other diffusion models (how to update model parameters by using feedback data) is still very challenging. If the offline influence function defined by a diffusion model based on a real application does not meet the monotonicity or submodularity, such as the complementary model \cite{guo2019novel}, how to analyze the regret bounds of its corresponding online problem? From a more general point of view, we wish to design an algorithm that can solve the Online IM problem without a specific diffusion propagation based on the CMAB framework, such as a data-driven method \cite{zuo2023dscom}, which is hard to get a theoretical bound until now.
    \item The CMAB framework that the Online IM problem is based on still has a lot of room for theoretical innovation. On one hand, the MAB problem is a special case of reinforcement learning (RL). The CMAB problem is a combination of combinatorial optimization and MAB problem, and the Offline IM problem has been successfully resolved by current RL-based approaches as shown in Section \ref{sec2-3}, thus it is hopeful to extend the Online IM problem to the generalized RL field. By combining with model deep learning techniques, we can explore a wider and more general framework of combinatorial online learning to cover more real application scenarios in the Online IM problem. On the other hand, it often has the characteristics of delayed feedback in practical application scenarios. For example in viral marketing, users are interested in a product under the influence of their friends, but they will not buy or promote it immediately, but will take practical actions at some point in the future. In the MAB problem, online algorithms and regret analysis under the delayed feedback, especially for multi-user delayed feedback \cite{li2023exp3}\cite{li2023adversarial}, have been provided. Back to the CMAB model, such a delayed feedback is worth considering and can be used for the Online IM problem.
    \item As we said in the second item, online algorithms are sensitive to different diffusion models, and then it will be more difficult to design online algorithms for diverse variant IM problems in real applications. In addition to the mentioned Topic-Aware IM \cite{chen2015online}, Budgeted IM \cite{nguyen2013budgeted}, and Competitive IM problem \cite{bharathi2007competitive} in Section \ref{sec5-2}, there are also a series of variants of the IM problem: Location-Aware IM \cite{cai2022survey}, Community-Aware IM \cite{guo2020influence}, Multi-Feature IM \cite{guo2020multi}, Signed IM \cite{yin2019signed}, and so on.
    First, according to current research, we find that they generally adopt contextual linear bandits to simulate the diverse backgrounds in the Online IM problem. However, this requires that the environment or background information can be encoded into a vector and this vector can form a linear relationship with parameters in the diffusion model, which is very difficult to implement in real applications. Second, for different variants in the Online IM problem, we need to design a specific framework and online algorithm, such as $\text{C}^2$MAB-T model for the Competitive IM problem, which has caused a huge workload. Up to now, due to the complexity and diversity of the problem, it is still impossible to form a unified framework with wide application scope. Third, at present, existing online algorithms are still limited in datasets, and most of the research work is carried out on artificial datasets, but not on real social datasets. How to apply our proposed Online IM algorithms to more real datasets and solve more practical problems is a research direction with important practical significance.
\end{itemize}

From the summarization and analysis presented, it is evident that addressing the Online IM problem in real-world social networks remains a complex challenge. Currently, our capabilities are largely confined to employing a basic CMAB framework within the IC model. This approach, however, has significant limitations and is not fully applicable to real-world scenarios. Its primary shortcoming lies in its inability to accommodate feedback that aligns with real-world situations and the variety of diffusion models encountered in practical applications. Furthermore, the foundational theory of combinatorial MAB itself offers substantial scope for enhancement. This situation opens up new avenues for research and invigorates the field. To effectively tackle these challenges, a comprehensive strategy is required, encompassing advancements in the fundamental theory of CMAB, the refinement of online algorithms and their theoretical underpinnings, and the modeling of real social network applications. A concerted effort in these three areas is essential for achieving meaningful progress.

\section{Conclusion}\label{sec7}
In this survey, we offer an extensive overview of the Online IM problem by covering both theoretical aspects and practical applications. For the integrity of this article and because the online algorithm takes an offline oracle as a subroutine, we first make a clear definition of the Offline IM problem and summarize those commonly used Offline IM algorithms, which include traditional approximation or heuristic algorithms and ML-based algorithms. Then, we give a standard definition of the Online IM problem and a basic CMAB framework, CMAB-T. Here, we summarize three types of feedback in the CMAB model and discuss in detail how to study the Online IM problem based on the CMAB-T model. This paves the way for solving the Online IM problem by using online learning methods. Furthermore, we have covered almost all Online IM algorithms up to now, focusing on characteristics and theoretical guarantees of online algorithms for different feedback types. Here, we elaborately explain their working principle and how to obtain regret bounds. Besides, we also collect plenty of innovative ideas about problem definition and algorithm designs and pioneering works for variants of the Online IM problem and their corresponding algorithms. Finally, we encapsulate current challenges and outline prospective research directions from four distinct perspectives.

In conclusion, online learning currently stands as a topic of significant interest. Its deployment in addressing online challenges within social networks holds vast potential for practical applications and commercial viability. It promises to address the issues inherent in traditional Offline IM research. This survey aims to provide researchers with a fresh perspective and serves as a comprehensive guide to the latest advancements and trends in the IM problem. It establishes a solid foundation for future research in this field.

\bibliography{ref}
\bibliographystyle{unsrtnat}

\end{document}

%% file: mypackages.tex
\usepackage[utf8]{inputenc}

\usepackage[T1]{fontenc}

\usepackage[numbers]{natbib} 

\usepackage{verbatim}
\usepackage[sumlimits,]{amsmath}
\usepackage{amsfonts}
\usepackage{amsthm}
\usepackage{amssymb}
\usepackage{srcltx}

\usepackage{tocloft}

\usepackage{bm}

\usepackage[bb=boondox]{mathalfa}   

\usepackage{array}
\usepackage{graphics,graphicx,psfrag,epsfig}
\usepackage{wrapfig}
\graphicspath{{.} {./figures/}}

\newcolumntype{M}[1]{>{\centering\arraybackslash}m{#1}}

\usepackage[parfill]{parskip}
\usepackage{geometry}

\usepackage{xcolor}
\usepackage{color}
\definecolor{deepblue}{rgb}{0,0,0.5}
\definecolor{deepred}{rgb}{0.6,0,0}
\definecolor{deepgreen}{rgb}{0,0.5,0}
\definecolor{deeporange}{rgb}{0.6,0.25,0}
\definecolor{verylightgray}{rgb}{0.97,0.97,0.97}
\definecolor{bluegray}{rgb}{0.3,0.3,0.6}

\usepackage[
    colorlinks=true,
    linkcolor=black,       urlcolor=bluegray,
    anchorcolor=bluegray,  filecolor=bluegray,
    menucolor=bluegray,    citecolor=bluegray
]{hyperref}

\usepackage{cleveref}
\usepackage{cancel}
\usepackage{ulem}
\newcommand{\stkout}[1]{\ifmmode\text{\sout{\ensuremath{#1}}}\else\sout{#1}\fi}

\DeclareFixedFont{\ttb}{T1}{txtt}{bx}{n}{8} 
\DeclareFixedFont{\ttm}{T1}{txtt}{m}{n}{8}  

\usepackage{listings}
\usepackage{textcomp}  

\lstdefinestyle{pythonstyle}{
  language=Python,
  basicstyle=\footnotesize\ttm,
  keywordstyle=\ttb\color{deepblue},
  commentstyle=\ttm\color{deepgreen},
  stringstyle=\color{deeporange},
  emphstyle=\ttb\color{deepred},
  backgroundcolor=\color{verylightgray},
  emph={__init__, __call__},      
  morekeywords = {self,yield},
  showspaces=false,               
  showstringspaces=false,         
  showtabs=false,                 
  frame=single,                   
  tabsize=2,                      
  captionpos=b,                   
  breaklines=false,               
  breakatwhitespace=false,        
  columns=fullflexible,           
  extendedchars=false,
  keepspaces=true,
  upquote=true,                   
  literate={-}{-}1 {*}{*}1
}

\lstnewenvironment{python}[1][]
{
\lstset{style=pythonstyle}
\lstset{#1}
}
{}

\newcommand\pythoninline[1]{{\lstset{style=pythonstyle}\lstinline!#1!}}


%% file: mymath.tex
\theoremstyle{plain}            
\newtheorem{thm}{Theorem}[section]

\theoremstyle{definition}       
\newtheorem{defn}[thm]{Definition}

\def\va{{\bm{a}}}

\def\vp{{\bm{p}}}

\def\vx{{\bm{x}}}